# Quantifying evolutionary dynamics of the basic genome of *E. coli*

Purushottam Dixit[1,2], Tin Yau Pang[2], F. William Studier, and Sergei Maslov
Biosciences Department, Brookhaven National Laboratory, Upton, NY 11973

**Abstract**  227 words

The ~4-Mbp basic genome shared by 32 independent isolates of *E. coli* representing considerable population diversity has been approximated by whole-genome multiple-alignment and computational filtering designed to remove mobile elements and highly variable regions. Single nucleotide polymorphisms (SNPs) in the 496 basic-genome pairs are identified and clonally inherited stretches are distinguished from those acquired by horizontal transfer (HT) by sharp discontinuities in SNP density. The six least diverged genome-pairs each have only one or two HT stretches, each occupying 42-115-kbp of basic genome and containing at least one gene cluster known to confer selective advantage. At higher divergences, the typical mosaic pattern of interspersed clonal and HT stretches across the entire basic genome are observed, including likely fragmented integrations across a restriction barrier. A simple model suggests that individual HT events are of the order of 10-kbp and are the chief contributor to genome divergence, bringing in almost 12 times more SNPs than point mutations. As a result of continuing horizontal transfer of such large segments, 400 out of the 496 strain-pairs beyond genomic divergence of $\Delta_c \sim 1.25\%$ share virtually no genomic material with their common ancestor. We conclude that the active and continuing horizontal transfer of moderately large genomic fragments is likely to be mediated primarily by a co-evolving population of phages that distribute random genome fragments throughout the population by generalized transduction, allowing efficient adaptation to environmental changes.

---

[1] Current address: Center for Computational Biology and Bioinformatics, Department of
[2] These authors contributed equally to this work

# Introduction

*E. coli* genomes are notoriously variable; containing an array of mostly phage-related mobile elements integrated at different sites, random transpositions of multiple transposable elements, and idiosyncratic genome rearrangements that include inversions, translocations, duplications, and deletions (1). Although *E. coli* grows by binary cell division, horizontal transfer of genome fragments among members of the evolving population has been recognized as an important factor in survival and evolution of the species (2, 3). Here, we use the term horizontal transfer specifically to mean acquisition of a genome segment from another member of the *E. coli* population by integration into the recipient genome by homologous recombination.

In contrast to point mutations (single nucleotide polymorphisms, SNPs), which can be harmful, neutral or advantageous, genome fragments acquired by horizontal transfer are usually from a well-functioning donor strain competitive for growth in the population and therefore likely to be neutral or advantageous (4–6). Horizontal transfer allows bacteria to sample diversity of the population, reduces clonal interference, and increases the rate of evolution (7).

The availability of complete genomes of *E. coli* strains covering a range of population diversity enables computational approaches to elucidate the relative contributions of random base-pair substitutions and horizontal transfer in divergence (2). We are interested particularly in the evolutionary dynamics of what we refer to as the "basic genome" of *E. coli*, the ~4-Mbp platform shared across the species, which excludes variable mobile elements and idiosyncratic genome rearrangements. As distinct from commonly derived ~2-Mbp core genome (2, 8), the appreciably larger basic genome derived here is not restricted to protein-coding sequence and has been aligned or rearranged to the genome order and orientation characteristic of the species-wide majority, facilitating direct comparisons between strains.

In this work, we first report a simple computational method to generate approximations to individual basic genome sequences from a multiple-alignment of whole genomes. We then quantify relative contributions of point mutations and horizontal transfers to genome-wide sequence variability. Surprisingly, we find that horizontally transferred DNA fragments are primarily in the size range of ~10-kbp to perhaps as large as ~250-kbp and that they replace a majority of the basic genome quite rapidly on the time scale of accumulation of SNPs by random mutation, quickly leading to a near-complete loss of clonally shared genome between pairs of strains. This leads us to conclude that horizontal transfers in *E. coli*, and probably many other bacterial species, are mediated primarily by a co-evolving population of bacteriophages that deliver genome fragments by generalized transduction.

# Results

## Deriving basic genome sequences from complete genomes

In order to derive the first approximation to the basic genome, through an iterative process, we settled on a set of 32 independently isolated *E. coli* strains from five broadly defined evolutionary groups A, B1, B2, D, and E (9). While the consensus organization of individual genomes is very similar to that of the comprehensively annotated K-12 laboratory strain MG1655, there were idiosyncratic rearrangements on some strains. In order to maximize the overlap between the strain-genomes, we modified some of the whole genomes prior to the multiple-alignment. We adopted the Mauve program (10, 11) for whole-genome multiple-alignment. See supplementary materials for details of the iterative process of strain selection, genome modification, and the multiple-alignment process. Phylogenetic relationships derived from the multiple-alignment are shown in Figure 1.

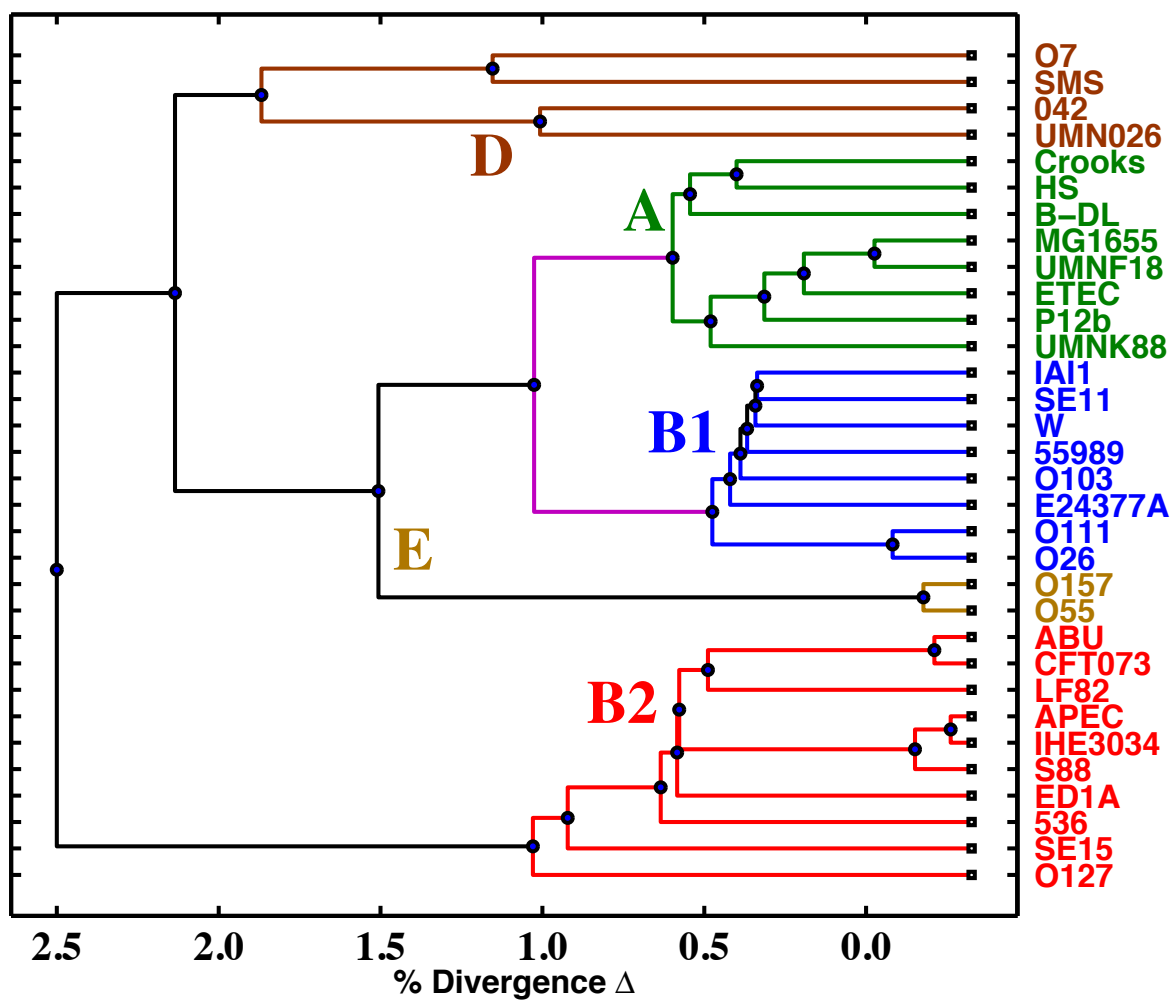

**Figure 1. The phylogenetic tree constructed from genomic divergence ∆ calculated from aligned basic genomes of 32 strains. Previously recognized ECOR evolutionary clades A (green), B1 (blue), B2 (red), D (dark brown), and E (light brown) are clustered together.**

The key step in deriving basic genomes from the output of the multiple-alignment was to apply a simple filter designed to eliminate mobile elements, idiosyncratic duplications, and parts or all of highly diverged regions that may be misaligned, while retaining group-specific and idiosyncratic deletions. The filter applied in the present analysis removed every base-pair position in which fewer than 22 of the 32 genomes have an aligned base pair. The requirement that the number of genomes represented be a minimum of 22 was chosen to be a point where addition of another genome causes minimal further reduction of basic genome length (See Figure 2). This filter reduced the initial 3044 Mauve alignment blocks to 105 blocks and initial ~22Mbp to 3,955,192 base pairs.

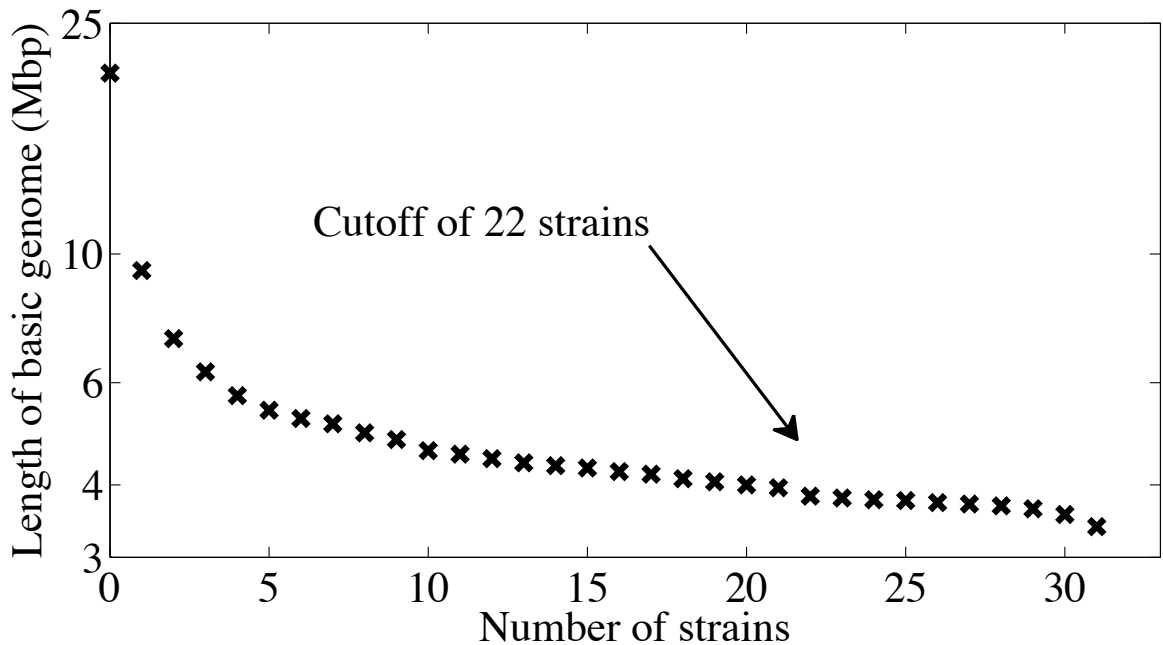

**Figure 2: The total length of the basic genome as a function of number of strains in the filter. The filter with *n* strains requires that a given base pair be included in the basic genome if and only if it is present in at least *n* of the 32 strains**.

The filter reduces individual basic genome lengths by variable amounts; the basic genomes are 68.7% to 84.4% of their corresponding complete genome lengths, reflecting the different percentages of individual whole genomes occupied by mobile elements. See supplementary materials for a detailed description of strain-specific comparison between the whole genome and the basic genome. In the supplementary data, we have deposited the basic genomes of all 32 strains.

## Genome segments for computational analysis

To study the distribution of SNPs on the basic genome, we divided the multiple-alignment in 1-kbp tandem segments regardless of gene-boundaries, 1-kbp being the typical length of a gene, other reasonable choices of segment length did not change our results (See supplementary materials for results with 500-bp and 2-kbp divisions). This way, the basic genome multiple-alignment was divided into 4008 1-kbp segments. Note that due to the applied filter (see above), individual strains potentially have less than 1-kbp in any given basic genome segment. We confirmed that on any given genome, a vast majority of segments (>85%) do indeed have all 1000 positions with a base pair. See supplementary materials for the statistics on segments.

50 out of these 4008 segments coded for tRNAs and ribosomal RNAs. These segments are known to undergo gene conversion i.e. is homologous recombination without horizontal transfer, giving rise to false SNPs (See supplementary materials for the location of these segments) (12). We excluded these segments from our analysis. In the rest of this paper, we work with 4008-50 = 3958 segments comprising ~3.91-Mbp of basic genome. While we do not differentiate between silent and non-silent SNPs, in the supplementary materials we show that silent SNPs contribute dominantly to the total number of SNPs and behave in a fashion very similar to the total number of SNPs. Our analysis below is insensitive to the distinction between the two.

## Evolutionary dynamics of basic genome variability

In a previous comparison of the genome sequences of the K-12 strain MG1655 and the reconstructed ancestral B genome referred to here as B-DL, we deduced a basic genome containing around 4000 protein-coding sequences (genes) with a total length of about 4-Mbp (13). We observed that SNPs are not randomly distributed among 3620 perfectly matched pairs of coding sequences in the basic genome but rather have two distinct regimes (13): 57% of the matched genes have sharply decreasing numbers of genes having 0, 1, 2 or 3 SNPs, well described by a Poisson distribution, with an abrupt transition to an exponential distribution of the remaining 43%, in which decreasing numbers of genes contain increasing numbers of SNPs from 4 to 102 SNPs per gene. The most prominent feature was the 35% of genes with no SNP at all. Genes in the two regimes of the SNP distribution are interspersed in clusters of variable lengths throughout the basic genome. We speculated that genes having 0-3 SNPs are primarily inherited clonally from the last common ancestor whereas genes comprising the exponential tail are primarily acquired by horizontal transfer from diverged members of the population.

The current study was undertaken to extend these results to *E. coli* population and to analyze how SNP distributions in the basic genome change as a function of evolutionary divergence between pairs of strains. Our primary measure of divergence between a pair of strains, referred to here as Δ, is the average percentage of SNPs across the entire alignment of the basic genomes of the two strains.

## Distribution of SNPs between pairs of basic genomes

In Figure 3, we plot the distribution of SNPs in 1-kbp segments for 4 genome pairs with divergence Δ = 0.5%-2.56%. Of the 4, the two least diverged pairs (Δ = 0.5% and 0.85%) show a SNP distribution similar to what we observed previously, namely, a peak of clonal segments containing 0-3 SNPs and an exponential tail of more highly diverged segments. 62 pairs out of 496 possible pairs of the 32 basic genomes have a similar pronounced clonal peak in the SNP distribution followed by an exponential tail (not shown, SNP table for all strains in supplementary materials). These include all 28 pairs within group A, all 28 pairs of B1 genomes, and the single pair of group E genomes. Only one of the 6 pairs of D strains and 4 of the 45 pairs of B2 genomes, which are more highly diverged groups, have a clonal peak. Except for the single D pair (Δ = 1.22%), all of these pairs have Δ less than 0.88%. See Table 1 for a summary of these 62 strains.

As divergence between pairs of genomes increases, the number of segments populating the clonal peak decreases, the average SNP density within the clonal segments increases, and the number of segments exponential tail increases while maintaining the same slope. None of the 388 pairs of genomes between different groups have a pronounced clonal peak. For the approximately two-thirds of genome pairs having Δ greater than ~1.5%, the distributions assume a broad maximum around 1% to 2% SNP density, and the decreasing numbers of segments having 0-3 SNPs are increasingly comprised of segments that specify highly conserved proteins, such as ribosomal proteins.

As we had speculated earlier, below, we will provide evidence and a simple computational model to support the claim that 1-kbp genome segments that have 0-3 SNPs primarily belong to the basic genome that is clonally inherited between the two compared strains (CC). On the other hand, 1-kbp segments that have 4 or more SNPs have been transferred horizontally in at least one of the two compared genomes (CR or RR). We now separately examine the dynamics of these clonally inherited and horizontally acquired genomic segments as a function of divergence Δ.

| Strain 1 | Strain 2 | Divergence Δ | Clonal fraction | $\langle\delta_{clonal}\rangle$ | $\langle\delta_r\rangle$ | $\langle l_{clonal}\rangle$ | $\langle l_r\rangle$ | $N_{clonal}$ |
|---|---|---|---|---|---|---|---|---|
| APEC | IHE3034 | 0.05 | 0.97 | 0.015 | 2.27 | 614.2 | 12.6 | 6 |
| ABU | CFT073 | 0.11 | 0.94 | 0.016 | 1.95 | 890.1 | 58.8 | 4 |
| O157 | O55 | 0.15 | 0.96 | 0.060 | 3.28 | 923.9 | 34.2 | 4 |
| APEC | S88 | 0.17 | 0.95 | 0.011 | 3.93 | 449.9 | 21.4 | 8 |
| IHE3034 | S88 | 0.17 | 0.94 | 0.013 | 3.63 | 1194.4 | 82.5 | 3 |
| O111 | O26 | 0.24 | 0.93 | 0.047 | 3.54 | 518.7 | 26.5 | 7 |
| K12 | UMNF18 | 0.30 | 0.88 | 0.027 | 2.41 | 426.6 | 48.7 | 8 |
| ETEC | K12 | 0.50 | 0.67 | 0.044 | 1.47 | 29.1 | 13.4 | 90 |
| ETEC | UMNF18 | 0.54 | 0.68 | 0.051 | 1.60 | 29.6 | 13.6 | 89 |
| K12 | P12b | 0.58 | 0.60 | 0.030 | 1.42 | 42.2 | 28.0 | 55 |
| P12b | UMNF18 | 0.63 | 0.59 | 0.033 | 1.51 | 53.1 | 35.4 | 43 |
| IAI1 | SE11 | 0.66 | 0.30 | 0.087 | 0.91 | 8.9 | 20.5 | 131 |
| IAI1 | W | 0.67 | 0.25 | 0.086 | 0.87 | 8.2 | 24.1 | 120 |
| SE11 | W | 0.67 | 0.26 | 0.081 | 0.88 | 8.6 | 24.1 | 118 |
| 55989 | IAI1 | 0.68 | 0.31 | 0.081 | 0.95 | 8.8 | 19.5 | 136 |
| 55989 | W | 0.70 | 0.25 | 0.086 | 0.90 | 8.1 | 24.1 | 120 |
| ETEC | P12b | 0.70 | 0.49 | 0.053 | 1.35 | 23.5 | 23.5 | 81 |
| O103 | W | 0.70 | 0.27 | 0.090 | 0.94 | 8.7 | 22.8 | 122 |
| 55989 | SE11 | 0.70 | 0.29 | 0.076 | 0.97 | 9.4 | 22.3 | 122 |
| O103 | O26 | 0.71 | 0.27 | 0.094 | 0.94 | 9.2 | 24.7 | 113 |
| IAI1 | O103 | 0.71 | 0.27 | 0.099 | 0.94 | 8.6 | 23.5 | 120 |
| O103 | SE11 | 0.72 | 0.25 | 0.097 | 0.94 | 8.5 | 24.4 | 117 |
| 55989 | O103 | 0.72 | 0.26 | 0.098 | 0.94 | 8.4 | 23.9 | 119 |
| Crooks | HS | 0.73 | 0.51 | 0.052 | 1.45 | 18.6 | 17.2 | 106 |
| E24377A | SE11 | 0.73 | 0.23 | 0.099 | 0.92 | 8.1 | 26.3 | 112 |
| E24377A | W | 0.73 | 0.24 | 0.094 | 0.93 | 8.1 | 25.3 | 116 |
| ETEC | UMNK88 | 0.73 | 0.42 | 0.067 | 1.22 | 11.7 | 15.8 | 140 |
| 55989 | O26 | 0.74 | 0.20 | 0.095 | 0.90 | 7.5 | 30.1 | 101 |
| K12 | UMNK88 | 0.74 | 0.43 | 0.051 | 1.28 | 12.1 | 15.4 | 140 |
| E24377A | IAI1 | 0.75 | 0.23 | 0.093 | 0.95 | 7.9 | 25.3 | 116 |
| 55989 | E24377A | 0.76 | 0.22 | 0.101 | 0.95 | 8.4 | 29.6 | 101 |
| O26 | W | 0.77 | 0.21 | 0.095 | 0.94 | 7.6 | 28.4 | 106 |
| IAI1 | O26 | 0.77 | 0.23 | 0.099 | 0.98 | 7.8 | 24.9 | 117 |
| E24377A | O103 | 0.77 | 0.24 | 0.097 | 0.99 | 8.4 | 25.5 | 113 |
| O103 | O111 | 0.78 | 0.27 | 0.099 | 1.03 | 9.0 | 24.0 | 116 |
| Crooks | P12b | 0.78 | 0.41 | 0.039 | 1.30 | 20.0 | 28.1 | 79 |
| O26 | SE11 | 0.78 | 0.20 | 0.092 | 0.96 | 8.0 | 31.3 | 97 |
| CFT073 | LF82 | 0.79 | 0.15 | 0.084 | 0.92 | 8.5 | 47.2 | 66 |
| E24377A | O26 | 0.80 | 0.20 | 0.101 | 0.97 | 7.9 | 30.7 | 99 |
| UMNF18 | UMNK88 | 0.80 | 0.43 | 0.057 | 1.37 | 12.0 | 15.5 | 140 |
| B-DL | ETEC | 0.81 | 0.35 | 0.071 | 1.22 | 10.6 | 18.9 | 129 |
| B-DL | Crooks | 0.83 | 0.36 | 0.061 | 1.28 | 10.5 | 17.8 | 134 |
| ABU | LF82 | 0.83 | 0.15 | 0.082 | 0.97 | 8.1 | 43.9 | 71 |
| 55989 | O111 | 0.84 | 0.19 | 0.102 | 1.02 | 7.5 | 30.8 | 99 |
| O111 | W | 0.85 | 0.20 | 0.100 | 1.04 | 7.6 | 29.1 | 104 |
| IAI1 | O111 | 0.85 | 0.24 | 0.108 | 1.08 | 7.9 | 25.0 | 117 |
| IHE3034 | LF82 | 0.85 | 0.09 | 0.103 | 0.92 | 6.6 | 64.0 | 53 |
| B-DL | MG1655 | 0.85 | 0.34 | 0.056 | 1.26 | 10.4 | 20.0 | 125 |
| O111 | SE11 | 0.85 | 0.19 | 0.100 | 1.03 | 8.2 | 34.2 | 90 |
| CFT073 | IHE3034 | 0.85 | 0.13 | 0.119 | 0.96 | 7.5 | 49.2 | 65 |
| APEC | LF82 | 0.86 | 0.09 | 0.106 | 0.94 | 6.7 | 68.1 | 50 |
| Crooks | ETEC | 0.87 | 0.31 | 0.070 | 1.24 | 9.6 | 20.7 | 126 |
| APEC | CFT073 | 0.87 | 0.11 | 0.126 | 0.97 | 7.0 | 53.2 | 61 |
| ABU | IHE3034 | 0.88 | 0.13 | 0.123 | 1.00 | 7.0 | 45.7 | 70 |
| B-DL | UMNK88 | 0.89 | 0.31 | 0.076 | 1.26 | 10.2 | 21.8 | 119 |
| ED1A | LF82 | 0.89 | 0.12 | 0.113 | 1.00 | 7.4 | 50.2 | 63 |
| ED1A | IHE3034 | 0.89 | 0.13 | 0.115 | 1.00 | 7.6 | 51.7 | 61 |
| E24377A | O111 | 0.89 | 0.20 | 0.115 | 1.08 | 8.0 | 32.1 | 95 |
| ABU | APEC | 0.89 | 0.12 | 0.130 | 1.00 | 6.8 | 48.3 | 67 |
| B-DL | P12b | 0.89 | 0.31 | 0.059 | 1.28 | 10.7 | 23.3 | 111 |
| APEC | ED1A | 0.90 | 0.13 | 0.119 | 1.01 | 7.6 | 50.8 | 62 |
| 42 | UMNF18 | 2.38 | 0.02 | 0.047 | 2.44 | 16.0 | 529.6 | 6 |
| | | | | | | | | |
| **Farthest *E. coli* strains** | | | | | | | | |
| O111 | O127 | 3.01 | | | | | | |
| | | | | | | | | |
| **Closest relatives** | | | | | | | | |
| K12 | *E. fergusonii* | 10.79 | | | | | | |
| K12 | *Salmonella Enterica* | 19.33 | | | | | | |

**Table 1: A summary of genome pairs whose SNP distribution shows a clear Poisson peak, indicating a significant fraction of genome that is clonally inherited. Notice that as genomic divergence ∆ increases, the SNP density in the clonal stretches also increases while the SNP density in the recombined stretches is somewhat constant. On the other hand, with increasing ∆, the length of clonal stretches decreases, the length of recombined stretches in creases, and the number of clonal stretches first increases and then decreases. To put the evolutionary divergences in perspective, we have also shown the maximum divergence between the studied strains, the divergence between K12 and *E. fergusonii* and *Salmonella Enterica.***

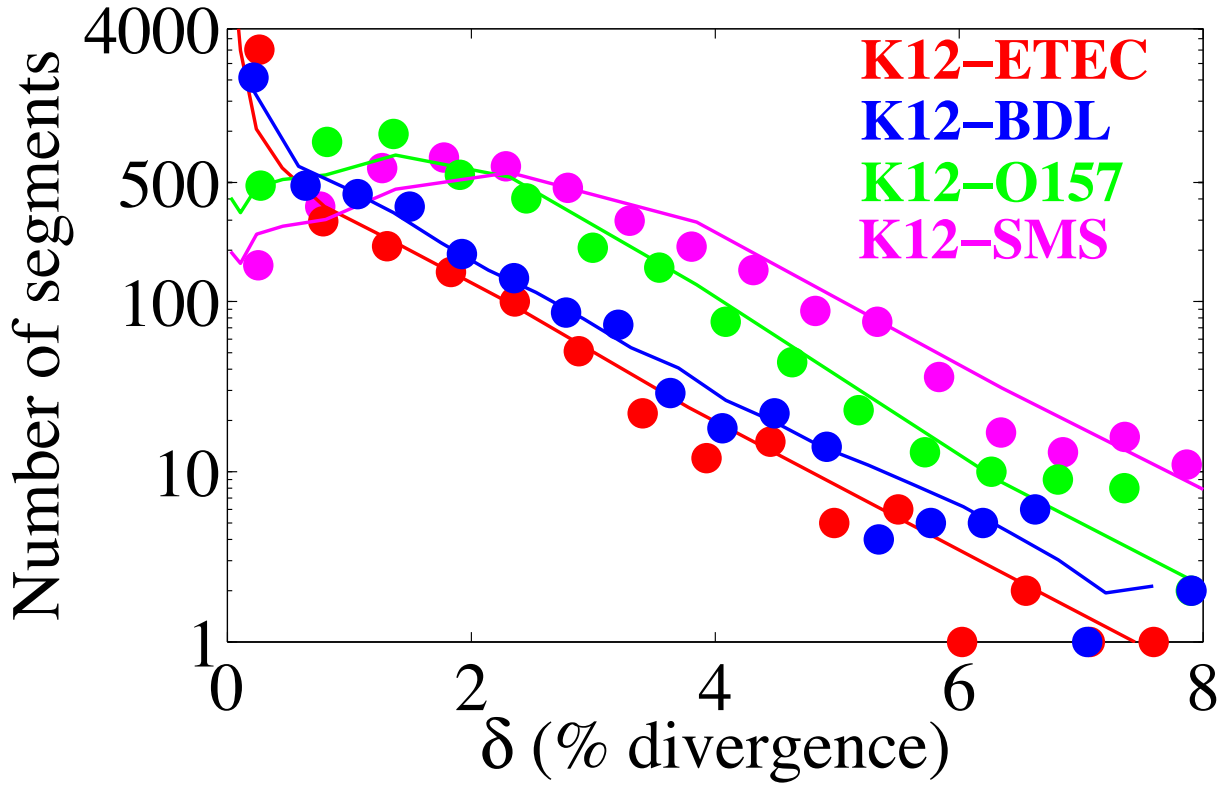

**Figure 3:** SNP distribution when comparing the basic genomes of strain K12-MH1655 with ETEC ($\Delta = 0.5\%$), BDL ($\Delta = 0.85\%$), O157 ($\Delta = 1.8\%$), and SMS ($\Delta = 2.56\%$). The global genomic divergence $\Delta$ between the compared pairs is shown in the Legend. $\delta$ denotes the %-divergence in 1-kbp segments of the basic genome (e.g. $\delta = 1\%$ is equivalent to 10 SNPs per 1-kbp). Notice the pronounced Poisson peak in the SNP distribution between K12 and ETEC and K12 and P12b. The Poisson peak consistently disappears when $\Delta > 0.88\%$. For highly diverged strains, the SNP distribution develops a broad peak between $\delta = 1\% - 2\%$ divergences. Right: a model fit for SNP distribution for genome-pairs K12 and ETEC and K12 and SMS. Even though the model parameters were chosen to fit population wide quantities (see below), the model independently fits the SNP distributions of two arbitrarily chosen pairs of genomes.

## Identifying clonal and recombined stretches on basic genomes.

For further analyses of clonally inherited and horizontally acquired SNPs, we made the assumption, based on patterns of overall SNP distributions, that 1-kbp segments containing 0, 1, 2, or 3 SNPs will almost always have been inherited clonally by both genomes since their last common ancestor (CC) whereas paired segments containing 4 or more SNPs will almost always have acquired them by homologous recombination resulting from horizontal transfer in either or both genomes (RC or RR).

We noticed that some segments naively classified as above as "clonal" may in fact be in the middle of large "recombined" stretches. Yet, the number of consecutive "clonal" segments within a stretch otherwise marked as "recombined" is usually less than 5 (see supplementary materials for statistics of segments wrongly classified as "clonal"). To address this mosaic nature of recombined stretches, from here onwards, we considered a given 1-kbp segment *clonal* between a pair of genomes *if* it was part of a contiguous stretch longer than 5-kbp of segments all with 0-3 SNPs. Segments which are not classified as *clonal* are classified as *horizontally transferred* from a co-evolving population. Changing the length of the contiguous stretch from 5-kbp does not alter our conclusions (see Supplementary materials for results with alternative cutoffs).

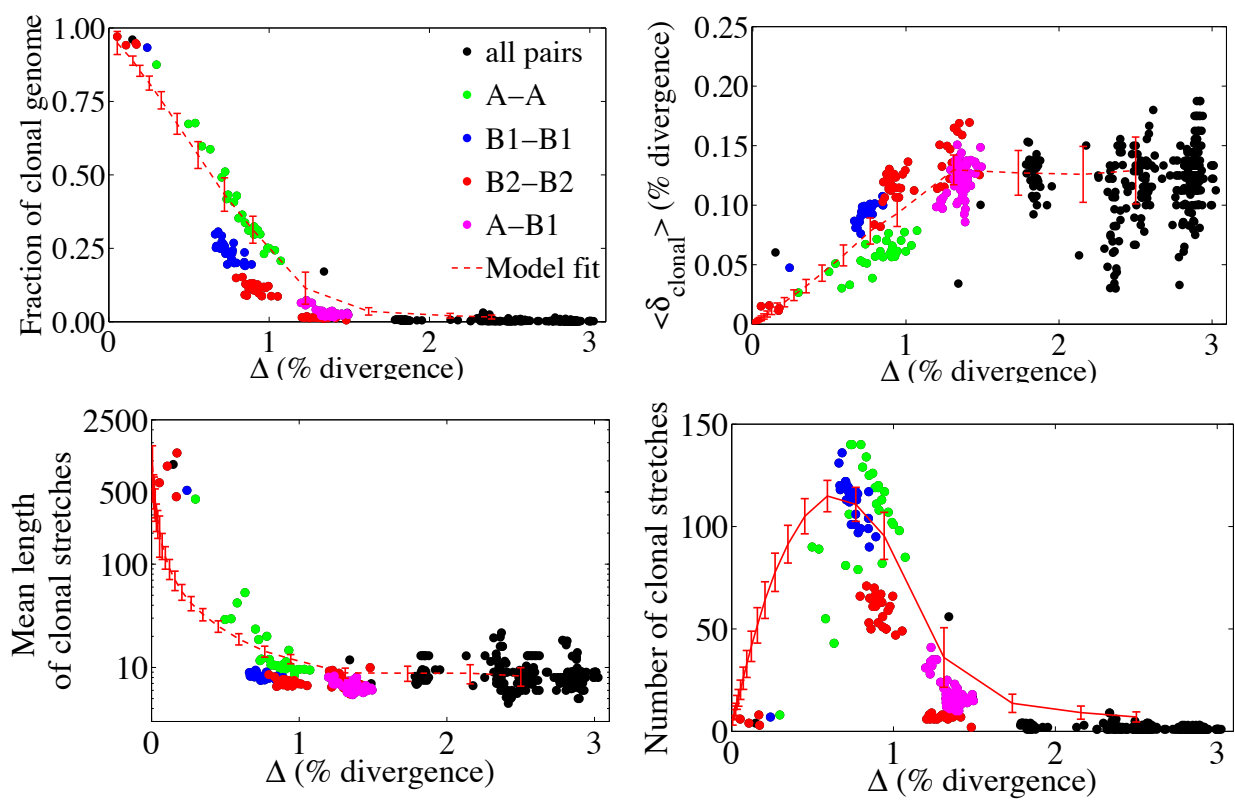

**Figure 4: Upper left: As a pair of genomes receives multiple horizontal transfers in the process of diversification, the fraction genome retained from the common ancestor in both the strains decreases, reaching ~20% around $\Delta = 0.9\%$. When genomes have diverged sufficiently, $\Delta > \Delta_c \sim 1.25\%$, the two strains share very little of the common ancestral genome owing to rampant horizontal transfer. Upper right: As a pair of strains diverges, the common ancestral stretches on their genomes acquire SNPs through random mutations. Consequently, the average SNP density in clonal stretches increases linearly with genomic divergence $\Delta$. When the common ancestral genome is nearly completely lost, there is no reliable way to detect it and the mean SNP density fluctuates. Lower left: Horizontal transfer causes the clonally inherited genome to be interspersed with genomic fragments from the population**

leading to a rapid decrease in the length of contiguous clonally inherited stretches. Beyond $\Delta = \Delta_c$, the mean length of clonal stretches falls to ~1-10 kbp. Lower right: As strains receive horizontal transfer, stretches of clonally inherited genome are broken up by horizontal transfer thus increasing the number of clonal stretches. Eventually, at $\Delta \sim 0.9\%$, independent horizontal transfer events start overlapping with each other and the number of clonal stretches no longer increases. When $\Delta > \Delta_c$ number of clonal stretches drops to 1-10. Red dashed line in all plots represents a model fit. The model parameters are mutation rate $\mu = 6.3 \times 10^{-10}$ per base pair per generation, recombination frequency $\rho = 1.7 \times 10^{-10}$ per base pair per generation, average length of horizontally transferred stretch $l_r = 7.5$-kbp, and finally "quality control" divergence $\delta_{qc} = 1\%$. See below for the description of the model. Note that model predictions in Figure 3B were also performed with the same parameters.

## Genomic segments inherited from last clonal ancestor in a pair of basic genomes

### The fraction of clonally inherited genome decreases linearly with increasing Δ and approaches 0 at $\Delta_c = 1.25\%$

If horizontal transfer of genomic segments from within the population happens at a constant rate with no strong dependence on the function of the genes involved, we expect the percentage of the basic genome remaining clonal between a pair of strains to decrease as time passes by. Indeed, the percentage of basic genome remaining clonal in both genomes (CC) in a pair of aligned genomes decreases linearly with increasing Δ to only ~20% of the paired genome length when Δ reaches ~0.9% (Figure 4A). Here we use Δ as a proxy for time, an excellent assumption if a vast majority of SNPs were acquired through horizontal transfers. Also, if the length of horizontally transferred fragments is large enough, beyond a critical divergence $\Delta = \Delta_c$, the clonally inherited genome fraction will fall very close to zero. Indeed, for strain pairs having Δ greater than $\Delta_c = 1.25\%$ (400 out of 496 strain comparisons), essentially all of the basic genome has been exchanged by horizontal transfer in one or both of the paired strains and the clonal percentage falls below ~5% to as low as ~1% of the paired genome length.

### The average SNP density in contiguous clonal stretches increases as a function of Δ and saturates at $\Delta_c = 1.25\%$

The average SNP density in clonal stretches reflects the rate at which point mutations occur on the basic genome. We observe that mean SNP density in clonal stretches increases linearly from 0.01% to 0.15% as Δ increases from 0.1% to 1.25%, indicating a constant rate of acquisition of SNPs through point mutations. Beyond $\Delta_c$, the number of clonal stretches is very small and comprises mainly of ribosomal proteins where SNPs are not likely to be neutral. Consequently, the average SNP density in the clonal regions fluctuates highly but does not appreciably increase as $\Delta$ increases.

## The average length of contiguous clonal stretches decreases as a function of $\Delta$ and approaches 0 at $\Delta_c = 1.25\%$

For a pair of very closely related strains, almost all of the basic genome is clonally inherited from the last common ancestor and the mean length of clonal stretch is very close to the total length of the basic genome. With time, as the basic genomes receive genomic fragments through horizontal transfer from the population, patches of recombined stretches intersperse the long stretches of clonally inherited genome thus decreasing the average length of contiguous clonal stretches. We observed that the average length of uninterrupted clonal stretches decreases monotonically from ~1000-kbp at $\Delta = 0.1\%$ to almost 1-10-kbp beyond $\Delta > 1.25\%$.

## The number of clonal stretches first increases as a function of $\Delta$ and falls of rapidly to near zero around $\Delta_c = 1.25\%$

For identical strains that have not experienced any horizontal transfer, the number of clonal stretches is simply one. As strains keep on receiving genomic fragments from the population, the horizontally transferred segments break the continuous stretches of clonal segments and as a result the number of clonal stretches increases as strains diverge while their length decreases (see above). If the length of the horizontally transferred fragments is large enough, at a certain divergence, the horizontally transferred segments start overlapping with each other thus decreasing the number of clonal stretches. When the genomes have diverged sufficiently, they are completely covered by transferred segments and the number of clonal stretches drops to zero as $\Delta$ approaches $\Delta_c = 1.25\%$.

# Genomic segments affected by horizontal transfer in a pair of aligned basic genomes

### Horizontal transfer in closely related strains may involve very long fragments:

Early horizontal transfer events can be easily detected in closely related strains by inspecting SNP distribution across 1-kbp segments. The six least diverged pairs of strains involve nine different genomes (Table 1). These basic genome pairs show a striking pattern: essentially all of the SNPs putatively acquired by horizontal transfer are in only one or two stretches of segments, each stretch extending across ~42-115 kbp of the basic genome sequence. The remaining minor fraction of putatively transferred segments (small isolated clusters of segments with 4 or more SNPs per 1kb) is dispersed across the clonal regions as individual segments or small clusters. The long stretches of recombinant segments are essentially continuous but are sometimes interrupted by short stretches of segments with 0-3 SNPs. These "clonal" interruptions reflect overlaps of clonal regions in the independent mosaic patterns of the donor and recipient genomes paired in the horizontally transferred region. See section Supplementary materials for statistics on this 'mosaic' pattern.

When we map the abovementioned long transferred segments on the individual strain genomes, each shows regions containing genes specifying O-antigens, DNA restriction and modification systems, type 1 fimbriae involved in adhesion, outer capsule proteins, etc.; all of these gene clusters have a clear selective advantage and were likely fixed early in divergence. Most of these exchangeable gene clusters occupy ~20-kbp in the complete genome sequence, some of which is highly variable and therefore eliminated by the filter used to derive the basic genome. Mapping the end-points of these long recombinant stretches to the complete genome sequence gives lengths of these horizontally transferred fragments of ~85-266 kbp. See supplementary materials for exact location of the early HT regions on the 9 basic genomes.

**Horizontal transfer and DNA restriction:**

Type I DNA restriction and modification systems are found in a significant fraction of *E. coli* strains and can initiate degradation of foreign DNA that does not have a protective modification (14). Our previous analysis of genome sequences of two *E. coli* B strains (13) detected two different patterns of integration in transduction by phage P1 across the type-I restriction barrier of B. A 1961 experiment (15) found that strains selected for a repair of 6-kbp maltose deletion in B by P1 transduction from K-12 strain W3110 not only repaired the maltose deletion but also introduced at least five additional fragments of minimum length 0.4-11.9 kbp. Overall, the six integrated fragments extended across at least 71.3 kbp implying that most of the ~100-kbp P1 phage is capable of transducing. A study by McKane and Milkman found that P1 transduction fragments integrated across restriction barriers in a wide range of independently isolated *E. coli* strains averaged 8-14 kbp (16).

Genome annotations indicate that O111 is the only one of the nine strains in the seven least diverged genome pairs to lack a type I restriction system. If so, why are essentially all of the horizontal transfers identified in these genome pairs larger are of the order of 40-120 kbp? It is highly unlikely that each of the minimum of 12 large horizontally transferred genome fragments came from a strain having the same restriction-modification specificity as the recipient.

The fragments that are typically integrated after transduction across an active type I restriction barrier are too small to import genomic fragments as large as ~40-120 kbp. However, when special situations (such as certain types of stresses) allow an entering genome fragment to become integrated essentially intact, a fragment that carries a gene cluster advantageous in that situation would initiate a competitive new lineage, fixing both the new fragment and whatever random mutations and mosaic pattern of horizontal transfers happened to be present in the genome that received it. Diverging lineages will independently accumulate both random mutations and random small fragments integrated across a restriction barrier, which will mostly be close to selectively neutral. Successive adaptations enabled by acquisition of highly advantageous gene clusters in large horizontally transferred fragments will fix whatever mosaic pattern is present in the recipient genome and also introduce the mosaic pattern of the newly acquired fragment.

**Signatures of horizontal transfer on genomes as a function of the divergence $\Delta$ between them:**

Steps in the continuing process of divergence of basic genomes by horizontal transfer are revealed in the six least diverged genome-pairs, each of which has only one or two horizontal transfers of advantageous gene clusters. Although an initiating event in clonal expansions seems likely to be acquisition of advantageous gene clusters by horizontal transfer of a large genome fragment, subsequent horizontal transfers in diverging lineages may well be due mostly to transfer across a restriction barrier, in which single events result in integration of one or more genome fragments of ~10 kbp spread across some fraction of the length of a much larger imported fragment.

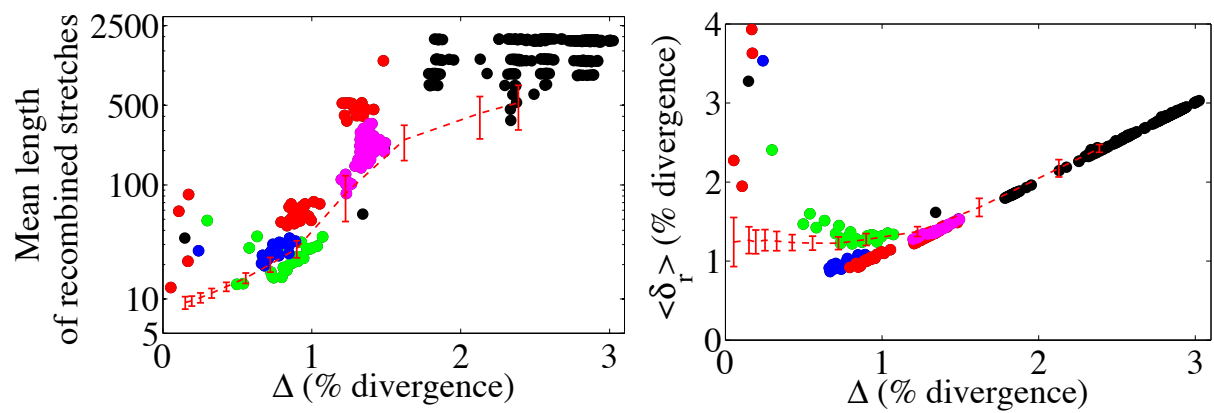

**Figure 5: Left: Average length of recombined stretches is roughly constant at lower divergence but when $\Delta > 0.9\%$, independent horizontal transfer events start overlapping with each other causing the average length of recombined stretches to increase rapidly. Beyond $\Delta > \Delta_c = 1.25\%$, nearly all of the genome is covered by recombination and the average length of recombined stretches is comparable to the basic genome length. Right: The density of SNPs in recombined stretches represents the population average of SNPs. While the 6 closely related genomes have peculiar transfer events (see Text), for genomes in $\Delta = 0.3\% - 0.9\%$, the average SNP density is weakly dependent on evolutionary group perhaps indicating differences in group-specific donor strains. When $\Delta > \Delta_c$, almost all of the genome is horizontally transferred and the density within horizontally transferred regions becomes equal to $\Delta$ itself. Green circles represent strain pairs within group A, blue within B1, and red within B2. Black circles represent all other strain comparisons.**

**The mean length of horizontally transferred stretches increases rapidly once independent horizontal transfer events overlap with each other**

For moderately diverged strains, distinct horizontal transfer events do not interfere with each other and the average length of horizontally transferred stretches reflects the mean length of incoming genomic fragments. Eventually, for genomes diverged beyond $\Delta > 0.9\%$, independent horizontal transfer events start overlapping with each other which results in a rapid increase in the average length of horizontally transferred stretches approaching the length of the entire basic genome as $\Delta$ approaches $\Delta_c = 1.25\%$.

**SNP density in recombined regions is weakly dependent on evolutionary groups**

Average SNP densities in recombined regions of pairs of basic genomes within different evolutionary groups cluster at slightly different levels over the range of $\Delta$ ~0.18-0.88% (Figure 5B). These differences might indicate that horizontal transfers in different groups derive from somewhat different groups of donor strains and/or that different groups have slightly different average levels of divergence from that of the entire population of *E. coli* genomes exchanging genome fragments. As $\Delta$ increases beyond the level where essentially all of the aligned length of paired basic genomes has been exchanged by horizontal transfer, the number of separable clonal stretches is rapidly reduced by overlaps of recombinant regions and $\Delta$ becomes equal to the overall SNP density of the paired genomes.

The high rates of horizontal transfer and attendant large fractions of recombinant SNPs can make accurate measures of clonal divergence difficult and uncertain. At low divergence, where >90% of basic genome length is clonal, average SNP densities in the long uninterrupted clonal stretches should give a relatively accurate measure of clonal divergence, which might be less than 10% the value of $\Delta$. In the intermediate range to $\Delta$ < 0.9%, where the longest stretches of uninterrupted clonal stretches may still be 50-100 kbp or more, the average SNP density of the longest clonal stretches should still be a reasonably accurate measure of divergence from the last common ancestor. However, beyond $\Delta_c$ ~ 1.25%, average SNP densities give some indication of relative divergence but are essentially meaningless as measures of clonal divergence times.

**Horizontal transfer in basic genomes is probably due primarily to transduction**

Our computational analyses of the basic genomes of this set of 32 independently isolated strains of *E. coli* indicate that horizontal transfer of genome fragments is ongoing continuously in nature, shuffling genome variability among members of the evolving population. Lineages diversify primarily by acquiring genome segments from diverged members of the population and completely replace their basic genomes before the diminishing genome fraction that has been clonally inherited accumulates an average SNP density as high as 0.2%. What is driving this process?

Well-studied mechanisms of genome exchange include conjugation, transformation and transduction. Generalized transduction by a co-evolving population of phages is by far the most likely mechanism to be responsible for the vast majority of horizontal transfer in the basic genome of *E. coli*. Phages are ubiquitous in nature and co-evolve with their bacterial hosts. They are known to be responsible for specific integration of most of the large variety of mobile elements that populate specific sites in complete bacterial genomes. The lysogenic phage P1 is a well studied example of a generalized transducing phage capable of packaging random fragments of genomic DNA as large as ~100 kbp without accompanying phage DNA and delivering them to any susceptible host. P1 itself would be capable of delivering many of the horizontally transferred basic-genome fragments we identify, and a variety of phages with similar capabilities are likely to be responsible for delivering genome fragments across the size range observed. Generalized transduction requires only co-evolution of the phage and bacterial populations and transfers random basic-genome fragments indiscriminately through the population for integration by homologous recombination. Regular transfer of fragments in the transducible size range may be in large part responsible for the ubiquitous operon structure in bacteria, which enables coordinated transfer of genes with different metabolic or operational functions.

*E. coli* is not known to have efficient mechanisms for uptake of DNA from its environment, present in some species of bacteria. Conjugation requires contact between cells and is probably not an efficient way to disseminate genome fragments through a dispersed population. Generalized transduction seems almost certain to be responsible for the vast majority of the horizontal transfer in *E. coli* and probably in many different bacterial species. Limits to the range of co-evolving populations of bacteria and generalized transducing phages may well be responsible for defining boundaries among different bacterial species.

# Concluding discussions

## Computational methods using basic genomes may be applicable for high-throughput analysis and annotation of the anticipated flood of bacterial genome sequences

The basic-genome framework has worked well for analyzing and interpreting evolutionary dynamics of *E. coli*. It readily identifies the default genome organization, idiosyncratic inversions, duplications and translocations; allows facile identification of group-specific features such as specific patterns of deletions or gene clusters; and highlights variable regions that have exchangeable versions evolved to be advantageous under different conditions. Types of mobile elements in the complete genome and their sites of integration in the basic genome are also readily identified and categorized.

The consensus basic genome derived here by an automated filter of a 32-genome Mauve multiple-alignment is an approximation that could be further refined to become a more comprehensive, fully annotated representation of the basic genome of *E. coli* and its characteristic variations. Integrating greater numbers of a wider variety of genomes would further define known and newly emerging subgroups. Databases of the different types of mobile elements represented throughout complete genomes of the species could be useful for automated annotations. Further refinement of computational methods might involve automated generation and refinement of consensus basic genome sequences for different bacterial species and their subgroups, which could be used to automate rapid categorization of the anticipated flood of new complete or draft genome sequences being produced by ever more efficient genome sequencing technology.

## A simple model explains essential features of the evolutionary dynamics of *E. coli*

Except for the early recombination events detected in closely related strains, we have studied the evolution of *E. coli* strains within the neutral framework viz. in the absence of selection. Recently, Fraser et al. (17, 18) proposed a simple neutral model that incorporates essential features of bacterial evolution. Here, we develop a modification of their work and analytically solve it. Briefly, the modified model involves $N$ (fixed) co-evolving bacterial strains $\{S_1, S_2, \ldots, S_N\}$ with $G$ genes each. The genes experience random mutations at a constant rate $\mu$ and every once in a while fragments of genome with average length $l_r$ are at a rate $\rho$ transferred between strains. An incoming segment with SNP density $\delta$ compared to the host genome is incorporated in the host genome with a probability exponentially decreasing in $\delta$ due to biophysical constraints of the homologous recombination process (19). We call this biophysical constraint "quality control". The "quality control" implies that an incoming segment with SNP density $\delta$ is accepted with probability $p \sim e^{-\delta/\delta_{qc}}$ where $\delta_{qc}$ is the quality control parameters. (See supplementary materials for details of the model)

Here, we are concerned with the dynamics of divergence between a pair of strains that evolve from a common ancestor. We analytically solve for the probability $P(\bar{\delta}(t) \to \bar{\delta}(t+1))$ where $\bar{\delta}(t) = \{\delta_1, \delta_2, \ldots, \delta_G\}$ is the collection of SNPs in aligned 1-kbp regions and $t$ is the number of generations since divergence between the two strains. We start the model with $\bar{\delta}(0) = \bar{0}$ (implying identical strains) and propagate a Markov chain for $\bar{\delta}(t)$ till the strains acquire a given genome wide divergence $\Delta = \langle \delta_i \rangle$.

We find that a model with the following parameters works fairly well: mutation rate $\mu = 6.3 \times 10^{-10}$ per base pair per generation, recombination frequency $\rho = 1.7 \times 10^{-10}$ per base pair per generation, average length of a horizontal transfer event $l_r = 7.5$-kbp, and finally the "quality control" exponential decay $\delta_{qc}$ to be 1% SNP divergence. The mutation rate, recombination frequency, and quality control exponential decay compare very well with previous independent estimates (2, 20, 21). The model predicts the observed SNP distributions viz. a clonal peak and an exponential tail at lower genome-wide divergence, which gradually changes to a peaked distribution (see Figure 3). It also predicts all the quantities plotted in Figure 4.

In the model, we precisely know which segments are clonally inherited in both genomes and which segments are recombined in at least one of them. Nevertheless, to be consistent with our analysis of real genomes, we employ the criterion used to defined clonal and horizontally transferred stretches in real genomes in model genomes as well. In supplementary materials we show that using the known history of segments in model genomes does not change our results much, providing further support for our definition of clonal and recombined stretches in real genomes. See supplementary materials for the details of the model, simulations, and fitting procedure.

**Recombination is the chief contributor to strain divergence**

When comparing two genomes, $r/\mu$ denotes the ratio of the SNPs that are brought about by recombination events to those that are brought about by point mutations. To a first approximation it is independent of evolutionary time $t$ elapsed since the last clonal ancestor of two strains. Indeed, assuming that mutations in the clonal segments are acquired at a constant rate $\mu$ (per base pair per generation) their density within clonal segments is given by $\langle \delta_C \rangle = 2\mu t$. On the other hand, recombination events within a given strain also happen at a constant rate $\rho$ per base pair per generation. Here we choose the convention assigning the start of the recombination event to the base pair closest to, say, the *oriC* site on the chromosome. Therefore, the fraction of the basic genome covered by recombined segment $f_r = 2\rho t l_r$. As seen in Figure 4X the density of SNPs within recombined segments $\langle \delta_r \rangle$ for $\Delta < \Delta_c = 1.25\%$, is approximately constant and equal to 1%. By combining all these factors one gets

$$r/\mu = \frac{2\rho t l_r \langle \delta_r \rangle}{2\mu t} = \frac{\rho l_r \langle \delta_r \rangle}{\mu}$$

which does not depend on evolutionary divergence time $t$. We estimate the average value of $r/\mu \approx 12$. Previously, Guttman and Dykhuizen (22) estimated that $r/\mu \approx 50$ by studying variations in short sequences of 12 *E. coli* strains. More recently, Touchon et al. (2) estimated $r/\mu \approx 1.5$. The value $r/\mu \approx 100$ derived in their paper is based on an alternative definition of $r/\mu$ that ignores whether a given nucleotide within a recombined segment is a SNP or not. Multiplying 100 by 1.5% which is the SNP density within recombined segments in their paper one gets the above mentioned $r/\mu \approx 1.5$. Our value $r/\mu \approx 12$ lies right in between these two earlier estimates.

**Existence of a critical divergence $\Delta_c \sim 1.25\%$**

Our analysis and the simple model both predict a major milestone in the evolutionary divergence of *E. coli* strains. As strains start diverging, they acquire point mutations and more importantly large chunks of horizontally transferred DNA from the population. Given that the average length of horizontally transferred segments is ~7.5-kbp, a pair of strains that have received of the order of hundreds of horizontal transfer events contain virtually none of their clonally ancestral genome. Our model predicts that this transition occurs around $\Delta_c = 1.25\%$.

Interestingly, a majority of genome pairs (400 out of 496) belong to this category ($\Delta > \Delta_c$), most comprising of genomes belonging to two different evolutionary clades. Thus, a majority of strains studied in this work no longer not share a common clonal ancestor with each other. If we define the *E. coli* species as a collection of strains that share a common clonally inherited genome from an ancestor, the species will be restricted to the individual clades. Clearly, a broader definition of species is required. While intra- and inter-species horizontal transfer has been considered important to bacterial evolution for some time (cite), we believe that we provide the first genome-wide and population-wide quantitative evidence towards to support the definition of *E. coli* as a sexual species.

**Exponential slope in SNP histogram is explained by a combination of population genetics and biophysics of homologous recombination**

The model predicts that the exponential slope in SNP histograms observed in Figure 3 may arise due to two effects viz. effective population size $N$ and the so-called "quality control".

Experiments on integration of foreign DNA into bacterial chromosome demonstrated that frequency of successful integration exponentially decays as a function of sequence divergence between the incoming DNA and the corresponding region of the host chromosome (19). Such decay has also been observed in wild meta-genomic samples (Banfield references). Biophysical mechanism of this decay can be understood as follows: homologous recombination requires two small continuous patches of $n$ mismatch-free nucleotides between the introduced DNA and the corresponding section of the host genome. The probability *p* of a successful recombination event is directly proportional to the number of such mismatch-free patches and falls of exponentially ($p \sim e^{-2n\delta}$) as a function of sequence divergence $\delta$ between the foreign DNA and the corresponding host genome.

This *quality control* mechanism that prevents foreign DNA from being integrated in the host genome also may acts as a mechanism shaping species boundary and could partially explain the exponential slope of $P(\delta)$ (see Figure 3). *If* the quality control mechanism is the sole contributor to the exponential slope i.e. if the effective population size $N$ is very large, we estimate that the length *n* of the patch required for homologous recombination to be $n \approx \frac{1}{2 \times 0.010} \approx 50$ base pairs, somewhat larger than the experimental estimate of 22 base pairs (19). Our estimate serves as the upper limit for length *n* of the mismatch-free patch required for homologous recombination.

On the other hand, in a fixed population of size $N$ the distribution of coalescence times *t*

(the number of generations to the last common ancestor) between two randomly selected members of the population follows the exponential function $exp(-t/N)$ (). Given the mutation rate $\mu$ (per base pair per generation) defining frequency at which neutral mutations are fixed in the population, genomic segments separated by $t$ generations would be characterized by SNP density $\delta = 2\mu t$. Hence the density of SNPs in gene-sized segments of two randomly selected members of the population will be exponentially distributed as $e^{-\frac{\delta}{2\mu N}}$. If effective population size is the sole determinant of the exponential slope, we estimate $N = (6.25 \pm 1.66) \times 10^8$. This estimate of the effective population size is a lower bound on the population size and agrees well with previous estimates (1). We suspect that in reality the exponential slope arises because of the combination of these two mechanisms.

## Acknowledgements
Work at Brookhaven was supported by grants PM-031 from the Office of Biological Research of the U.S. Department of Energy.